\documentclass[aps,pre,reprint,superscriptaddress,amsmath,amssymb,nofootinbib]{revtex4-2}

\usepackage{xcolor}

\usepackage{comment}

\usepackage{graphicx}
\usepackage{dcolumn}
\usepackage{bm}

\begin{document}

\title{Motional Degrees of Freedom in Network Hamiltonian Models}

\author{Peng Huang}
\affiliation{Department of Sociology, University of Georgia, Athens, Georgia 30602, USA}
\author{Elizabeth M. Diessner}
\affiliation{Department of Chemistry, University of California at Irvine, Irvine, California 92697, USA}
\author{Carter T. Butts}
\affiliation{Departments of Sociology, Statistics, Computer Science, and EECS, University of California at Irvine, Irvine, California 92697, USA\\ Corresponding author: \texttt{buttsc@uci.edu}}

\date{August 21, 2026}

\begin{abstract}
Network Hamiltonian Models (NHMs) provide an efficient framework for modeling the aggregation of interacting particles (e.g., the condensation of proteins into gel-like, oligomeric, or fibrillar states), representing the system as a network whose edges represent bound interactions.  Terms within the network Hamiltonian represent multi-body interactions governing aggregation behavior, and are specified via topological degrees of freedom.  Because motional degrees of freedom are not explicitly represented within the NHM, their influence must be indirectly accounted for by introduction of corresponding terms to the Hamiltonian.  Here, we describe specifications for terms representing motional degrees of freedom of two, three, and four-body interactions.  We also consider the impact of these terms on the aggregation states of a minimal system governed only by a pairwise edge potential, showing that three and four-body interactions favor condensation of the system into small droplet-like structures at low temperature.
\end{abstract}

\maketitle

\section{Introduction}

Widely used for modeling social systems (e.g., \cite{goodreau.et.al:d:2009,steglich.et.al:sm:2010,schaefer:sn:2012,krivitsky.morris:aas:2017,thomas.et.al:pnas:2020}), network-based models of interaction and structure formation have also proven useful for modeling physical systems (with examples ranging from quantum information and entanglement networks \cite{zaman.lee:prE:2019,poteshman.et.al:prr:2023} to studies of the properties of biomolecular aggregates and complex materials \cite{steinhardt.et.al:prB:1973,matthews.et.al:prE:2006,cheianov.et.al:prl:2007,diessner.et.al:jctc:2024}).  Among such approaches, network Hamiltonian models (NHMs) have shown promise as a tool for coarse-grained modeling of molecular systems \cite{grazioli.et.al:jpcB:2019}.  NHMs represent such systems as networks of bound interactions among particles, with the behavior of the system governed by an effective energy function (the eponymous network Hamiltonian).  Formally equivalent to the widely studied exponential-family random graph models (ERGMs) \cite{holland.leinhardt:jasa:1981,lusher.et.al:bk:2012}, NHMs are able to exploit from the latter a large and growing body of computational and statistical methodology \cite{hunter.et.al:jcgs:2012,schweinberger.et.al:ss:2020}, as well as theory gleaned from studies of ERGM behavior in multiple domains \cite{haggstrom.jonasson:jap:1999,park.newman:prevE:2004,butts:sm:2011b,schweinberger.handcock:jrssB:2015,butts:jms:2021}.  To date, NHMs have been successfully used to model such phenomena as protein fibrillization \cite{grazioli.et.al:jpcB:2019,yu.et.al:nsr:2020}, the formation of gel-like, unstructured, and oligomeric macromolecular aggregates \cite{diessner.et.al:jpcB:2023,diessner.et.al:jctc:2024}, and transient structure in intrinsically disordered proteins \cite{grazioli.et.al:fmb:2019}, where their ease of scaling to large systems on long timescales makes them a useful complement to other modeling techniques.

Unlike many other coarse-graining approaches, NHMs do not directly represent motional degrees of freedom; these are typically treated as time-scale separated with respect to the formation and dissolution of bound interactions, and are not necessarily coupled to the interaction network.  This approximation is not always sufficient, however.  For instance, bond vibrations among pairs of interacting particles will contribute to the total energy of the aggregation graph, an effect that drives bond breaking at high temperatures; to capture such effects arising from unmodeled motional degrees of freedom, it is necessary to approximate them via additional terms to the network Hamiltonian.  Two-body effects from bond vibrations can be accounted for by a term related to the number of edges in the network of bound interactions \cite{grazioli.et.al:jpcB:2019}, but higher-order effects have not previously been considered.

In this paper, we introduce NHM effects to approximate unmodeled motional degrees of freedom from three-body (libration) and four-body (dihedral oscillation) interactions, in the time-scale separated regime.  As we show, these effects can be expressed in terms of simple graph statistics, which can be easily employed within an NHM/ERGM framework.  To probe the effect of these motional terms on NHM behavior, we simulate the behavior of a minimal system of aggregating particles driven by a base interaction potential plus varying orders of approximation to the motional energy.  As we show, incorporating higher-order terms changes the aggregation behavior of the system at low temperature, inducing a transition to a phase of droplet-like aggregates with limited interaction.

The remainder of the paper is structured as follows.  Section~\ref{sec:nhm} briefly reviews the NHM formalism, and related kinetic extensions.  Section~\ref{sec:dof} introduces terms for approximating the energy from unmodeled motional degrees of freedom, and the respective orders of approximation (from no correction through four-body interactions).  Section~\ref{sec:sim} examines the impact of these corrections on model behavior via simulation of aggregation in a simplified particle system, and Section~\ref{sec:conc} concludes the paper.

\section{Network Hamiltonian Models} \label{sec:nhm}

The NHM represents a system of interacting particles, such as macromolecules or peptides in solution, as a network with nodes representing individual particles and edges defined by binary values indicating bound interactions between particle pairs. Here, we briefly review the treatment of NHMs in equilibrium, followed by a simple model of NHM kinetics; both are employed in the development that follows.

Consider a system of $N$ distinguishable particles, each pair of which may be in a ``bound'' or ``unbound'' state, in thermal contact with a reservoir at constant temperature $T$.  We represent the microstates of this system via an \emph{aggregation graph,} $G$, whose nodes represent individual particles and whose edges represent bound interactions.  The equilibrium behavior of $G$ may be specified in terms of an (effective) network Hamiltonian, $\mathcal{H}$, representing the energy of the system, and an entropic correction (reference measure) $h$, representing the effects of unmodeled degrees of freedom on the entropy of the aggregation graph.  Specifically, an NHM represents the equilibrium distribution of the aggregation graph at temperature $T$ by the Boltzmann distribution
\begin{equation}
    \label{eq:bolz}
    \Pr(G=g | \mathcal{H}, T) = \frac{\exp\left( -\mathcal{H}(g)/(k_B T) + \ln h(g)\right)}{Z(\mathcal{H},T)}
\end{equation}
\noindent where $k_B$ is the Boltzmann constant. The distribution is normalized by $Z(\mathcal{H},T)=\sum_{g'\in \mathbb{G}}\exp[-\mathcal{H}(g')/(k_B T) + \log(h(g'))]$, the partition function with respect to the ensemble of possible graph microstates $\mathbb{G}$.

$\mathcal{H}$ is usually written via the linear form $\mathcal{H}(g)= \phi^\intercal t(g)$, where $t:\mathbb{G} \to \mathbb{R}^p$ represent topological degrees of freedom of the aggregation graph and $\phi \in \mathbb{R}^p$ represent the energy per unit change in the corresponding degrees of freedom. Elements of $t$ may be flexibly chosen to represent potentially complex multi-body forces, and frequently involve counts of specific subgraphs or induced subgraphs of $G$ (as illustrated in Figure~\ref{fig:concept}A); the simplest element of $t$ is the edge count, $t_e$, which corresponds to the number of edges in its argument.  Although complex choices of $t$ are possible (see e.g., \cite{pattison.robins:sm:2002,snijders.et.al:sm:2006}), our interest here is in a small number of terms arising from motional degrees of freedom (as described below).  The log of the reference measure $h: \mathbb{G} \to [0,\infty)$ accounts for the entropic contribution of unmodeled degrees of freedom, of which the most important for aggregation graphs is the need for collisions to occur in order to form bound interactions.  Following \cite{grazioli.et.al:jpcB:2019}, we here employ the Krivitsky reference \cite{krivitsky.et.al:statm:2011} $h(g)=N^{-t_e(g)}$, which corrects for this constraint for a system at constant concentration \cite{butts:jms:2022}; as $N$ is here fixed, the system is thus implicitly in the $N,V,T$ ensemble.

We observe that the equilibrium distribution of $G$ under an NHM with $\mathcal{H}$ defined with respect to basis $t$ has an exponential family random graph (ERGM) form \cite{schweinberger.et.al:ss:2020},
\begin{equation}\label{eq:ergm}
    \Pr(G=g|t,\theta)=\frac{h(g)\exp(\theta^\intercal t(g))}{\sum_{g'\in \mathbb{G}}h(g')\exp(\theta^\intercal t(g'))}
\end{equation}
\noindent where the topological degrees of freedom $t$ here play the role of sufficient statistics, and model parameters $\theta \in \mathbb{R}^p$ are related to energy parameters $\phi$ by the relation $\theta^\intercal t(g) = -\phi^\intercal t(g) / (k_B T)$.  ERGMs have been widely studied as statistical models for relational data \cite{holland.leinhardt:jasa:1981,wasserman.pattison:p:1996,schweinberger.et.al:ss:2020}, and tools for simulation of graph distributions in ERGM form are well-established \cite{hunter.et.al:jss:2008,wang.et.al:sw:2009,krivitsky.et.al:jss:2023}; we exploit this connection in our simulation studies below.

Although the above formulation does not specify the kinetics of transitions between graph states, \cite{grazioli.et.al:jpcB:2019,yu.et.al:nsr:2020} proposed an ERGM generating process (EGP) \cite{butts:jms:2024} based on local Arrhenius kinetics for this purpose; this EGP is formally equivalent to the longitudinal ERGM (LERGM) introduced in earlier work by \cite{koskinen.snijders:jspi:2007} for modeling social network dynamics.  In its NHM formulation, the LERGM EGP is a continuous-time stochastic process in which graph $G$ in state $g_i$ transitions to distinct state $g_j$ at rate
\begin{equation}
r_{ij} = \begin{cases} \frac{A}{1+\exp{[\beta\Delta_{ij}^\mathcal{H}-\Delta_{ij}^h]}} & \text{if }g_j \in \mathcal{N}(g_i) \\ 0 & \text{otherwise} \end{cases}, \label{eq_lergm_rate}
\end{equation}
\noindent where $A$ is a pacing constant (collision rate), $\mathcal{N}(g_i)$ is the set of Hamming neighbors of $g_i$ in $\mathbb{G}$,  $\beta=1/(k_BT)$, and $\Delta_{ij}^\mathcal{H}=\mathcal{H}(g_j)-\mathcal{H}(g_i)$ and $\Delta_{ij}^h=\log(h(g_i))-\log(h(g_j))$ are the respective changes in energy and graph state entropy in moving from $g_i$ to $g_j$. The Hamming metric constraint ensures that bond formation/breaking is non-simultaneous, while the transition rates follow an approximate Arrhenius law when $\Delta_{ij}^\mathcal{H}-\Delta_{ij}^h \gg 0$ and $\Delta_{ij}^h\approx 0$.  It follows from the above that the exit rate from $G=g_i$ is given by $R_i=\sum_{j\in \mathcal{N}(g_i)}r_{ij}$, and the dwell time in state $g_i$ is exponentially distributed with expectation $R_i^{-1}$.  We employ this below in probing state stability.

NHMs have shown promise as simple and statistically tractable models for amyloid formation \cite{grazioli.et.al:jpcB:2019,yu.et.al:nsr:2020}, formation of protein gels and oligomeric states \cite{diessner.et.al:jctc:2024}, unstructured protein aggregates \cite{diessner.et.al:jpcB:2023}, and (when employed at the residue level) for conformational ensembles of both folded \cite{yin.butts:plos:2022} and intrinsically disordered \cite{grazioli.et.al:fmb:2019} proteins.  Although most work to date has focused on application to interaction among protein monomers (or residues thereof), they have potential utility as coarse-grained models for a wide range of particle systems.  Different systems may motivate different choices of $t$, the topological degrees of freedom for the effective Hamiltonian.  Here, we consider terms arising from unmodeled motional degrees of freedom, which are present in many typical settings.  

\section{NHM Terms for Motional Degrees of Freedom}\label{sec:dof}

In representing the aggregation state of our system entirely via its topology, we implicitly marginalize across motional degrees of freedom; for processes such as protein fibrillization in which bound interactions between monomers form slowly compared to the timescale of molecular motion, this can be a reasonable approximation.  However, motional degrees of freedom nevertheless contribute to energy of the system, and those that are directly coupled to the state of the aggregation graph will contribute to the network Hamiltonian.  Specifically, we must consider energetic contributions from motional degrees of freedom that are constrained by the topology of $G$.  Under the assumption of time-scale separation, each such degree of freedom can be approximated as being in equilibrium with respect to $G$, and thus (by the equipartition theorem) contributes $1/2 k_B T$ to the total energy.  It then remains to count the relevant degrees of freedom, which will be functions of the graph state.

In what follows, we will assume that (apart from $G$), particles are free to move in three dimensions.  We note some modifications for systems constrained to a two-dimensional surface, though we do not treat this problem in depth here.

\begin{figure*}
	\includegraphics[width=\textwidth]{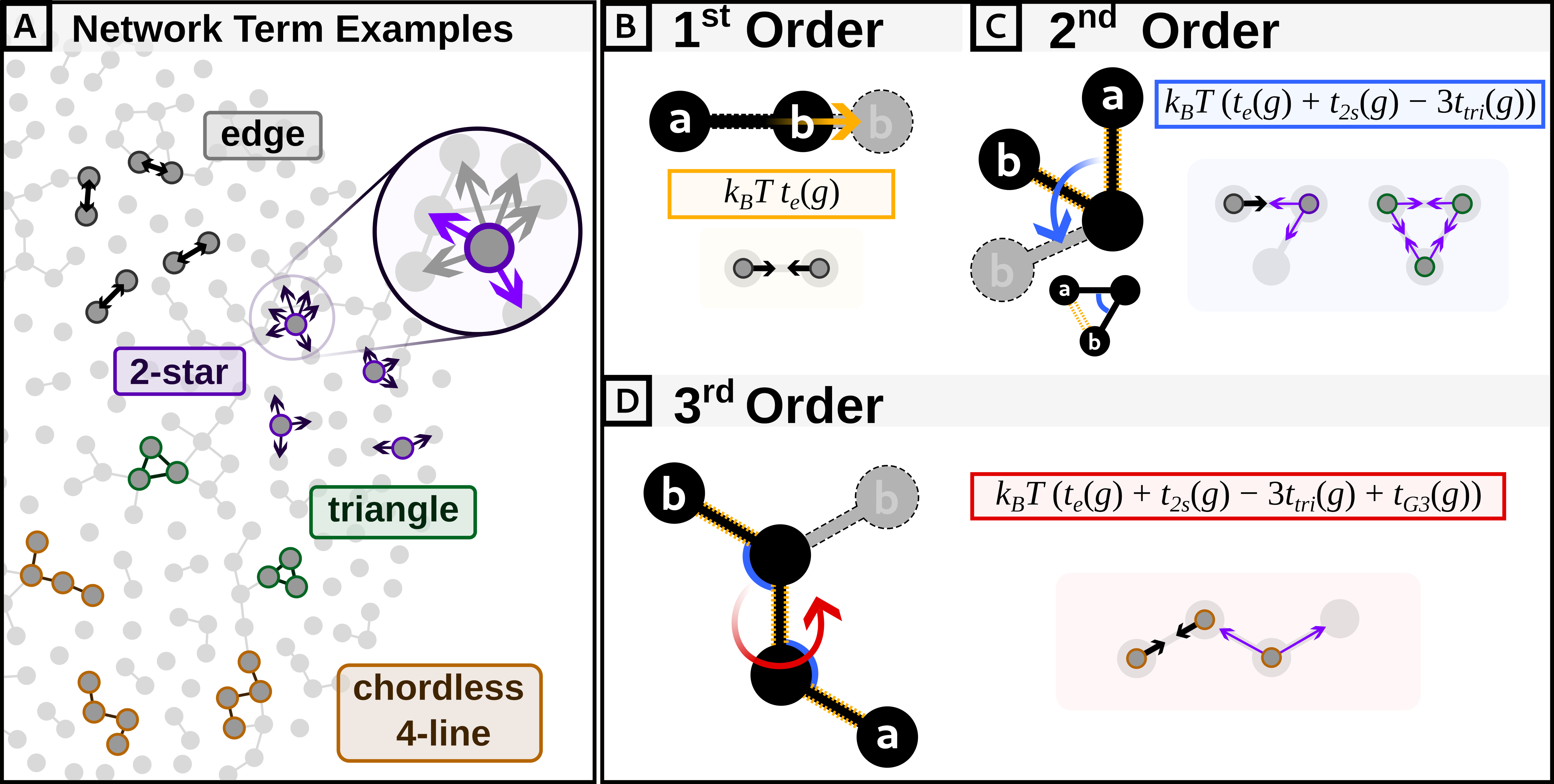}
	\caption{\label{fig:concept} Network terms represent topological degrees of freedom of the aggregation graph, and are captured as parameters in the network Hamiltonian. A) Examples of structures contributing to terms used here on an arbitrary graph state, including edges (grey), 2-stars (purple), triangles (green), and chordless 4-lines (orange). A close-up of a 2-star shows one of the $\tbinom{6}{2}$ 2-stars centered on a single node; every pair of edges on the same node contributes a 2-star.  B) First-order effects arise from bond vibrations between bound particles `a' and `b', indicated by a yellow arrow, contributing 2 df per edge ($t_e(G)$).  C) Second-order effects arise from the angular oscillations (librations) between particles `a' and `b' bound to a common partner, with the angle indicated by a blue arrow. These contribute an additional 2 df per open 2-path ($t_{2s}(G)-3t_{tri}(G)$). D) Third-order effects arise from oscillations of the dihedral angle between the planes containing particles `a' and `b' and their respective first and second-order neighbors (dihedral angle indicated by a red arrow).  These contribute an additional 2 df per chordless 4-line ($t_{G3}(G)$).}
\end{figure*}

The most basic example of a topology-constrained motion is the movement of two bound particles vis-\`a-vis each other (i.e., bond vibrations); see Figure~\ref{fig:concept}B.  In the classical regime, we can view each edge of $G$ as an oscillator with two degrees of freedom (kinetic and potential), contributing a total of $2 t_e(G)$ motional degrees of freedom to the graph.  Assuming that these motions are in equilibrium relative to the graph state then contributes a total energy of $k_B T t_e(G)$ to $\mathcal{H}(G)$.  We refer to this as the \emph{first-order correction,} reflecting energy from two-body interactions.  Note that the effect of this term is to penalize edge formation, with edges becoming increasingly unfavorable as $T\to \infty$; without this effect (as observed by \cite{grazioli.et.al:jpcB:2019}), the NHM shows unphysical behavior at high temperatures.

Where two particles are jointly bound to a third, it is possible for there to be a preferred angle between the two bonds; in this case, the angular oscillations between the bonds (librations) provide another contribution to the motional energy coupled to $G$.  This is illustrated in Figure~\ref{fig:concept}C.  We observe that every 2-star (i.e., every $i,j,k$ subgraph such that $i$ is bound to $j$ and $k$) contributes one such angular constraint, and hence two coupled degrees of freedom.  However, if $i,j,k$ form a 3-clique (i.e., a triangle), all three angles are already constrained by the respective bond lengths, and these degrees of freedom are hence redundant with those associated with bond vibrations.  Observing that every triangle contains three 2-stars, the total count of bond angle degrees of freedom is thus $2(t_{2s}(G) - 3 t_{tri}(G))$, where $t_{2s}$ and $t_{tri}$ are respectively the 2-star and triangle count statistics; this is equivalent to twice the number of open two-paths.  Assuming time-scale separation then gives us a contribution of $k_BT(t_{2s}(G) - 3 t_{tri}(G))$ from three-body interactions, net of the motional contribution from two-body interactions.  We refer to this as the second-order correction to $\mathcal{H}$ for motional degrees of freedom.  We observe that the effect of this correction is to disfavor open two-paths.

The third-order correction (Figure~\ref{fig:concept}D) considers motions associated with four-body interactions.  Most such degrees of freedom are already accounted for by two and three-body interactions, with the sole remaining exception being dihedral angles arising in chordless four-lines (i.e., induced subgraphs $i,j,k,l$ forming a path, with no other edges present).  For systems with preferred dihedral angles, each such configuration will contribute $2 t_{G3}(G)$ coupled degrees of freedom (with $t_{G3}$ referring to the chordless four-line as graphlet 3 in the nomenclature of \cite{milenkovic.przulj:ci:2008}).  Under time-scale separation, this adds $k_B T t_{G3}(G)$ to $\mathcal{H}$, disfavoring the formation of extended structures that are unbuttressed by triangles.  Note that, however, chordless four-lines can arise in partially triangulated environments.  For instance, if two disjoint cliques of orders $n_1,n_2$ are bridged by a single edge, this will create $(n_1-1)(n_2-1)$ chordless four-lines.  This implies that another effect of dihedral oscillations is to disfavor local bridging of densely clustered (hence compact) aggregates.  As we show below, this can have a substantial behavior on aggregate stability.

As the third-order correction shows, higher-order motions are increasingly constrained by lower-order ones; further, it is not evident that many systems will have preferred motional behavior involving five or more bodies.  Thus, we restrict ourselves here to corrections of these first three orders.  Although, as noted, we focus on the three-dimensional case, we also observe that dihedral angles cannot be realized in two dimensions, and thus only the first two corrections can be expected to be relevant in that setting.  By turns, only the first-order correction (bond vibrations) is applicable in the one-dimensional case.  

Given these corrections to the network Hamiltonian for time-scale separated motional degrees of freedom, we now investigate their effects on aggregation.  For this, we consider a minimal system containing only an edge potential and motional degrees of freedom, allowing us to consider the impact of the latter in isolation.

\section{Simulation Study} \label{sec:sim}

To probe the effect of motional degrees of freedom terms on NHM aggregation graphs, we simulate the behavior of minimal NHMs containing only an edge potential (i.e., a general propensity for bond formation) and corrections for motional degrees of freedom.  As we show, motional energy terms can be consequential for aggregation behavior, particularly at low temperature. 

\subsection{Research Design}

We begin by defining an NHM whose Hamiltonian has a single term (the edge count, $t_e$), reflecting a baseline edge potential.  Without loss of generality, we choose our units such that $k_B=1$ and the base dissociation energy of an edge is one unit (i.e., $\phi_e=-1$).  We then consider extensions of the base model incorporating first, second, or third-order corrections for motional degrees of freedom (respectively).  For all models we use the Krivitsky reference measure $h(G)=N^{-t_e(G)}$, along with the packing constraint that the support ($\mathbb{G}$) consists of all order-$N$ graphs with maximum degree $\le 12$; these respectively reflect the entropic correction for particles needing to be co-located for bond formation to occur\cite{grazioli.et.al:jpcB:2019,butts:jms:2019} and a physical constraint on coordination number (using a model of sphere-like packing).  The Hamiltonians for the four models examined here are summarized in Table~\ref{tab:setup}; note that all models fall within the general class of exponential family distributions with sufficient statistics $t_e,t_{2s},t_{tri},$ and $t_{G3}$ (counts of edges, two-stars, triangles, and chordless four-lines).  Denoting the degree-constrained graphs by $\mathbb{G}_{d\le 12}$, the NHMs with the above choices of units have the graph distribution
\begin{multline}
    \Pr(G=g | \mathcal{H}, T) = \\
    \frac{\exp( -\mathcal{H}(g)/T  - (\log N) t_e(g))}{\sum_{g' \in \mathbb{G}_{d\le 12}} \exp( -\mathcal{H}(g')/T  - (\log N) t_e(g'))}. \label{eq:fullergm}
\end{multline}

\begin{table*}
    \centering
    \caption{NHM terms and Hamiltonians for the baseline and motionally corrected models (units chosen so that $k_B=1$); effects indicate the correction added at each order.}
    \begin{tabular}{clll}
    \hline
       Order  & Effects  Added & Sufficient Statistics & \multicolumn{1}{c}{$\mathcal{H}(G)$} \\ \hline
       0  & Edge Potential        & $t_e$                       & $-t_e(G)$ \\
       1  & Bond Vibrations       & $t_e$                       & $(T-1) t_e(G)$ \\
       2  & Librations            & $t_e,t_{2s},t_{tri}$        & $(T-1) t_e(G) + T(t_{2s}(G) - 3 t_{tri}(G))$\\
       3  & Dihedral Oscillations~ & $t_e,t_{2s},t_{tri},t_{G3}$ & $(T-1) t_e(G) + T(t_{2s}(G) - 3 t_{tri}(G) + t_{G3}(G))$ \\\hline
    \end{tabular}
    \label{tab:setup}
\end{table*}

We can see immediately from Table~\ref{tab:setup} that the order 0 and order 1 models behave as degree-constrained (homogeneous) random graphs, while higher-order models have more complex structure.  We thus probe model behavior by simulating properties of network structure for each model as a function of temperature.  We employ a network with $N=500$ nodes, varying temperature from $T=0.1$ to $T=10$ in an exponentiated arithmetic sequence.  (Beyond $T=10$, networks under all models are sufficiently sparse that there is little contribution from higher-order terms, and the models show similar behavior.)  For each model at each temperature we simulate 500 draws from the equilibrium graph distribution using Markov chain Monte Carlo (MCMC), specifically employing the default tie/random-dyad algorithm in the \texttt{statnet} \textbf{R} package \cite{handcock.et.al:jss:2008,hunter.et.al:jss:2008}, uniformly thinned with 2.5$\times 10^{7}$ discarded thinning draws per draw retained, following 2.5$\times 10^7$ discarded burn-in draws. 
    
For each simulated network, we compute nine summary statistics describing different aspects of the aggregation state.  These are: (1) \emph{density}, the proportion of node pairs that are adjacent; (2) \emph{components,} the number of maximal connected node sets (i.e., distinct aggregates); (3) \emph{monomers,} the number of isolated monomers (nodes with no edges); (4) \emph{dimers,} the number of isolated dimers (i.e., node pairs that are only connected to each other); (5) \emph{mean component size}, i.e. average number of nodes per component; (6) \emph{two stars/vertex}, the number of $i,j,k$ subgraphs in which $j$ is adjacent to $i$ and $k$ divided by $N$; (7) \emph{triangles/vertex,} the number of 3-cliques divided by $N$; (8) \emph{transitivity,} the proportion of two paths that are closed ($i-j-k \Rightarrow i-k$); and (9) \emph{unbuttressed edges,} the number of edges that have no shared partners.  We examine the behavior of these properties as a function of temperature and model order below.

While the zero and first-order models have smooth behavior in the low-temperature regime, we observe more complex transitions for the second and third-order models at approximately $0.12\le T \le 0.13$.  To probe this regime, we perform an additional set of simulations for these two models, again taking 500 draws at each temperature with $N=500$.  Temperature was varied from $2^{-4}$ to $2^{-2}$, with MCMC used at each temperature to obtain draws.  As above, 2.5$\times 10^7$ burn-in iterations were discarded for each Markov chain, with a thinning interval of 2.5$\times 10^7$ for each subsequent draw.  Examples of typical realizations were also retained for visualization.

To gain additional insights into kinetics in the low temperature regime, we simulate trajectories using the kinetic model of Eq.~\eqref{eq_lergm_rate}, beginning with a sample of 500 free monomers and proceeding until the condensed state is reached; this was observed to occur within 4$\times 10^5$ transitions.  We also directly calculate the change rates in homogeneous phases of $k$-cliques of order from 1 to 13 (the maximum possible, given the coordination constraint); this was performed on systems of size $N=1000$, with events involving excess vertices (for $N$ not exactly divisible by $k$) removed for purposes of rate calculation.  Trajectory simulation and hazard calculations were performed using the \texttt{ergmgp} package \cite{butts:sw:2023}.

\subsection{Equilibrium Behavior by Temperature} \label{sec:eqsim}

\begin{figure*}[t]
\includegraphics[width=\textwidth]{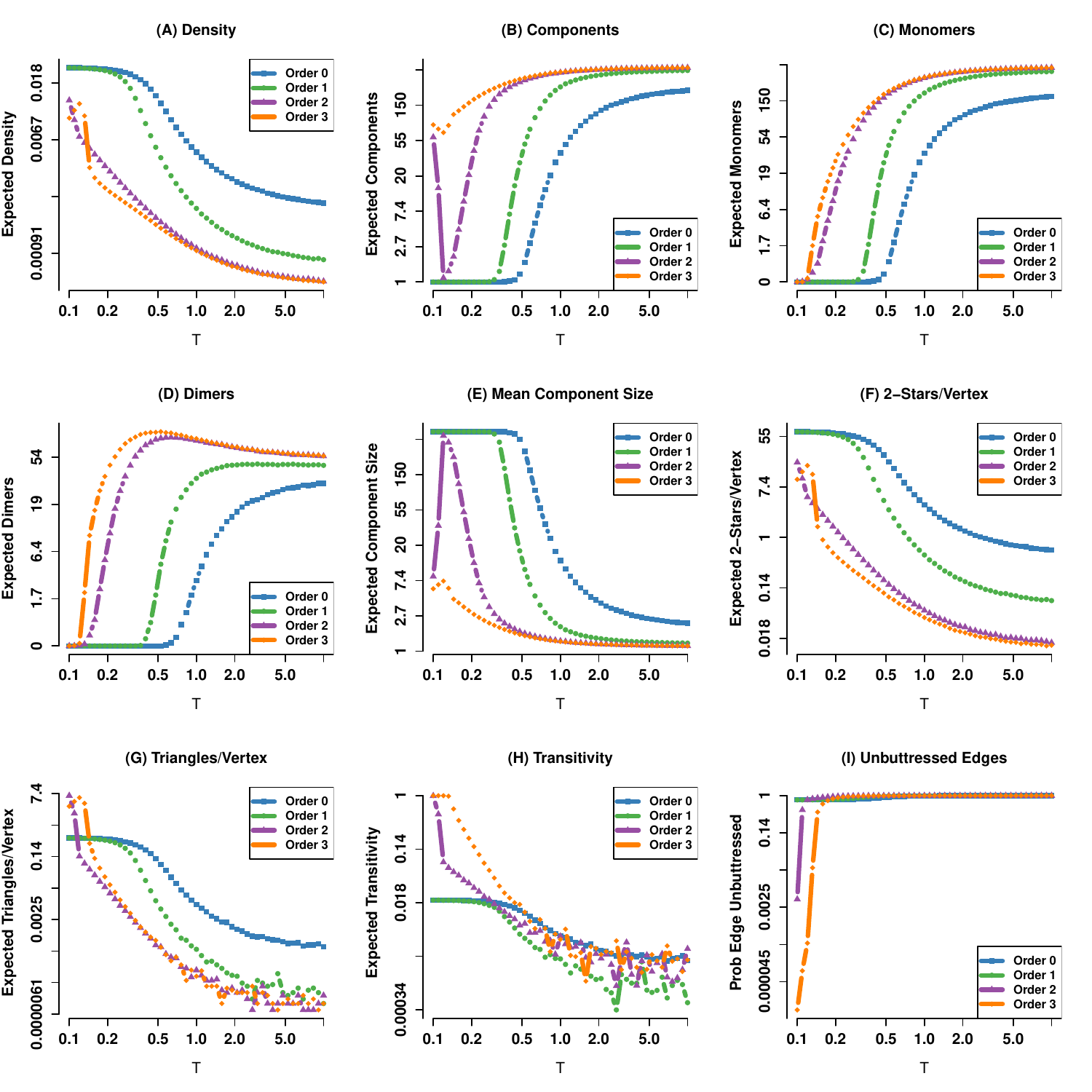}
\caption{\label{fig:rslt1} Mean aggregation graph properties (y-axes) as a function of temperature (x-axis) by model order (color).  Motional corrections substantially affect aggregation states, particularly at low temperature.}
\end{figure*}

%
Figure~\ref{fig:rslt1} shows the mean properties of the equilibrium aggregation graph as a function of temperature, by model.  As observed, the 0-order (baseline) and 1st-order (bond vibration) models behave as Bernoulli graphs when the expected mean degree is substantially lower than the packing constraint ($\bar{d}\ll 12$), which occurs when $T\gg 0.4$ for the 0-order model and $T\gg 0.29$ for the 1st-order model.  At substantially lower temperatures, these models approach uniform degree-regular graphs, with degree equal to the packing constraint for both models; at high temperatures, by contrast, the effect of the motional correction is to reduce the expected degree by a factor of approximately $e\approx 2.7$.  (We make use here of the fact that, for a Bernoulli graph under the Krivitsky reference, $\bar{d}\approx \exp(\theta_e)$ for $N$ large, where $\theta_e$ is the edge parameter in the model's ERGM representation.)  Physically, this reflects the reduced stability of bonds at high temperature due to vibrational motion.  It may be observed that the limiting mean degree as $T\to\infty$ in the 0-order model is $\approx 1$, implying that this model retains considerable edge structure in the high-temperature limit.  The corresponding mean degree limit in the bond vibration model is $1/e\approx 0.37$, which is considerably more physical: as seen from the component and monomer counts (Panels B,C), relatively little high-temperature structure remains once the effect of vibrational energy is accounted for.  That said, a small fraction of dimers and higher-order oligomers remain in this regime.  By turns, increasing density at low temperature leads to the usual transition to a well-connected regime seen in uniform random graph models \cite{bollobas:bk:2001}; there is little local closure (low transitivity), leading to a locally loose, gel-like state for both models.

Incorporation of libration and dihedral oscillations substantially alters this picture.  By penalizing open 2-paths, both the 2nd and 3rd-order models destabilize non-triangulated oligomers at high temperature; since triangles have little chance of forming in this low-density regime, however, this is not their primary effect.  Instead, we note that adding an edge between nodes $i$ and $j$ with respective degrees $d_i,d_j$ will add, in the absence of triangulation, $d_i+d_j$ 2-paths to the graph.  In the sparse regime, the librational cost of an edge addition increases with the degree of the endpoints, thus penalizing high-degree nodes (and connections among them).  At constant density, this has the effect of reducing the degree variance by spreading edges as equally as possible. At low density, this favors formation of independent dimers, followed by predominantly linear oligomers as density increases.  While triangles are not energetically disfavored, they continue to be \emph{entropically} disfavored, since low density and uniform low maximum degree provide relatively few opportunities for them to arise.  Adding the dihedral term has little additional effect in this regime, since most components are too small to produce chordless 4-lines at high temperature.  As temperature falls and density increases, we begin to see divergence between the two models, with the dihedral term disfavoring long linear structures; where networks generated by the second-order model start to coalesce into larger components, we see the third-order model instead forms small, linear oligomers of highly regular size (since longer oligomers produce more 4-lines).

\begin{figure*}[t]
\includegraphics[width=\textwidth]{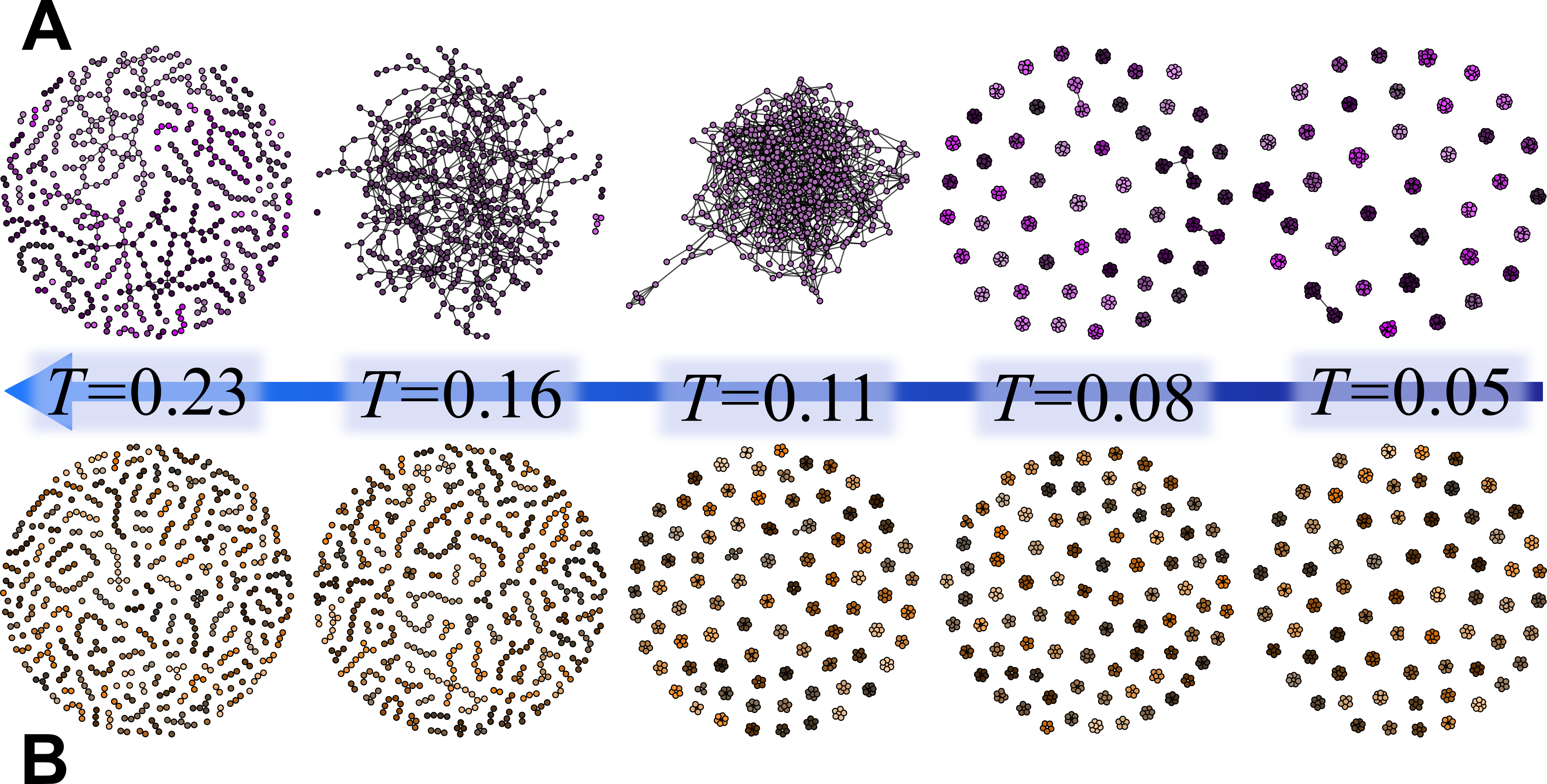}%
\caption{\label{fig:coldexamp} Typical aggregation graphs at extreme low temperature for 2nd (A) and 3rd (B) order models.  Nodes are shaded to show distinct components.  Both models transition from minimally aggregated states at $T=0.25$ to compact particles by $T=0.0625$, but show different behavior at intermediate temperatures.}
\end{figure*}

\begin{figure*}
\includegraphics[width=\textwidth]{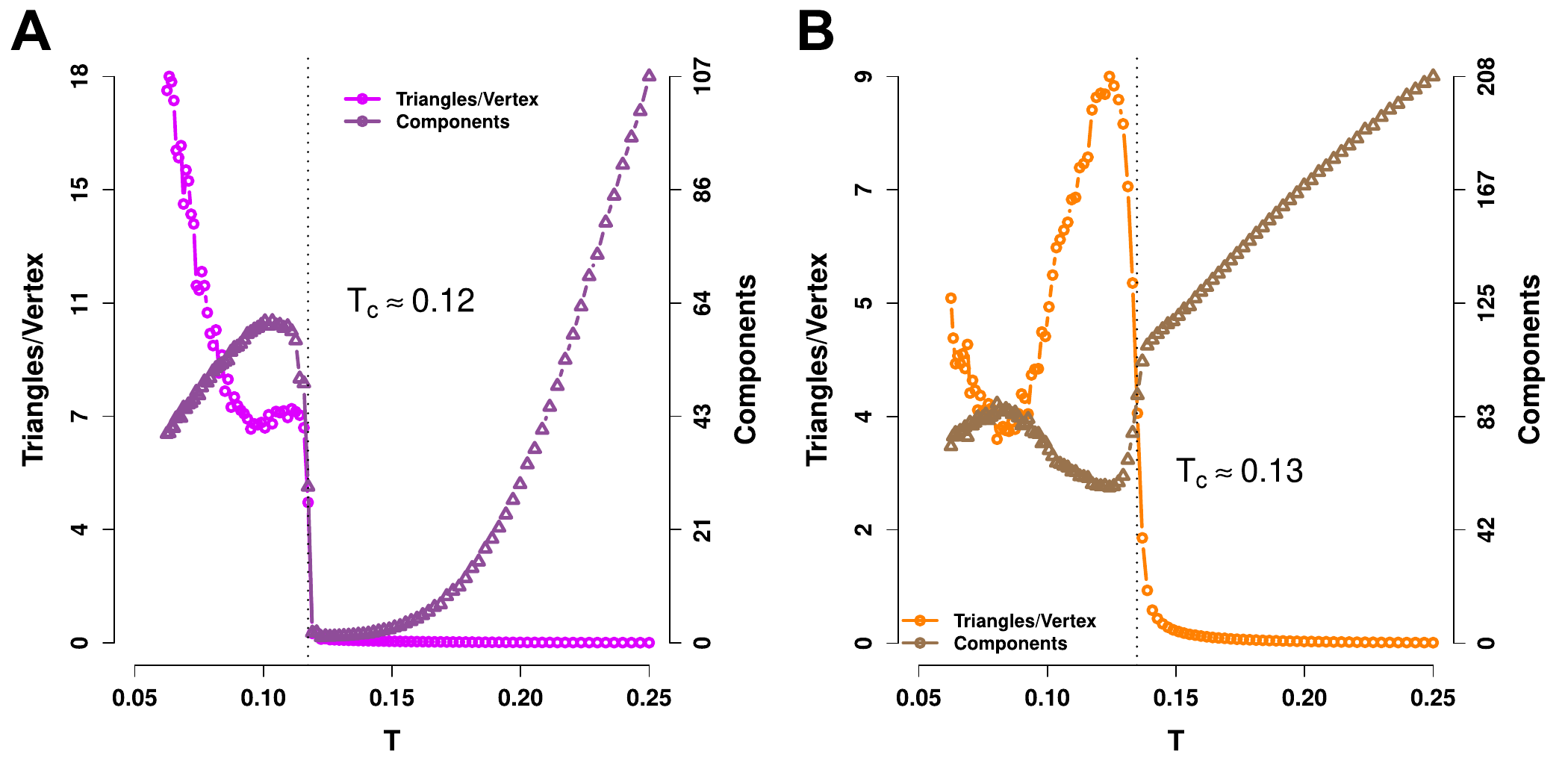}
\caption{\label{fig:coldtrans} Phase behavior of 2nd (A) and 3rd (B) order models near the first low-temperature transition.  Neither model generates triangulation above $T_c$, but adding a dihedral effect locally reverses the behavior of component count across the transition.
}
\end{figure*}

Turning to the low-temperature regime, we see greater exaggeration of the differences between the models with and without dihedral oscillations.  Figure~\ref{fig:coldexamp} shows typical realizations from both models at low temperatures.  As noted above, aggregation begins in both models with formation of dimers and short linear oligomers; highly branched structures are disfavored due to the cost of libration energy from forming high-degree positions in the absence of triangulation.  In the 2nd-order model, these largely linear structures give way at lower temperatures first to larger dendritic structures, and then to a loose gel phase when the mean degree passes the connectivity threshold.  This gel phase is still largely untriangulated, as the underlying graphs are sparse and the ``piling up'' of edges into triangles is entropically disfavored.  Continued cooling leads eventually to a second transition (at $T_c \approx 0.12$), to a phase of dense, compact aggregates that are largely disconnected from each other (Figure~\ref{fig:coldtrans}A).  Although this droplet-like phase is entropically unfavorable (there are relatively few droplet-like graphs, relative to gel-like graphs), it allows much larger numbers of edges to be sustained without producing open two-paths.  The cost of librational energy thus drives condensation at low temperature.

By contrast, addition of dihedral oscillations disfavors elongated structures, preventing the growth of the large dendritic aggregates and gel condensation seen in the 2nd-order model.  One instead sees an increasing prevalence of short linear oligomers of very similar size (since longer oligomers are systematically disfavored), followed by a sharp transition to a phase dominated by dense, compact clusters at $T_c\approx 0.13$ (Figures~\ref{fig:coldexamp}B~and~\ref{fig:coldtrans}B).  Unlike the 2nd-order case, these aggregates remain consistently well-separated, due to the large number of chordless 4-lines created by bridging pairs of otherwise disjoint clusters.  The clusters themselves are able to sustain higher coordination numbers (degrees) than in the sparse oligomeric phase due to the relative stability of closed versus open two-paths; as edges become more favorable at very low temperature, it is energetically favored to concentrate them in cliques rather than to place them in open structures.

Taken together, we thus observe that motional degrees of freedom tend to act as structure breakers at high temperature, with vibrational effects reducing density, librational effects regularizing degree, and dihedral effects regularizing component size.  At low temperatures, the bond vibration model converges to the baseline (zero-order) model, both of which produce gel-like states.  By contrast, librations in the absence of dihedral effects lead first to a sparse, untriangulated network of dendritic and largely linear oligomers, before transitioning first to a connected gel phase and then a collapsed phase of compact, droplet-like dense oligomers with occasional contact.  Adding dihedral effects prevents the formation of the large oligomers and hence the gel phase, with a single transition to the compact droplet phase.  These effects are driven primarily by (1) the impact of bond vibrations on density, (2) the suppression by librations of open two-paths, and (3) the suppression of dihedral oscillations of long, chordless linear structures, along with (4) the entropic cost of ``protecting'' two-paths by condensing edges into triangles.  Although this last loses out below the critical temperature, it is decisive above it (as evidenced by the extremely low rate of triangulation above the transition to the droplet phase).

\subsection{Model Kinetics at Low Temperatures}\label{sec:clique}

\begin{figure*}[t]
\includegraphics[width=\textwidth]{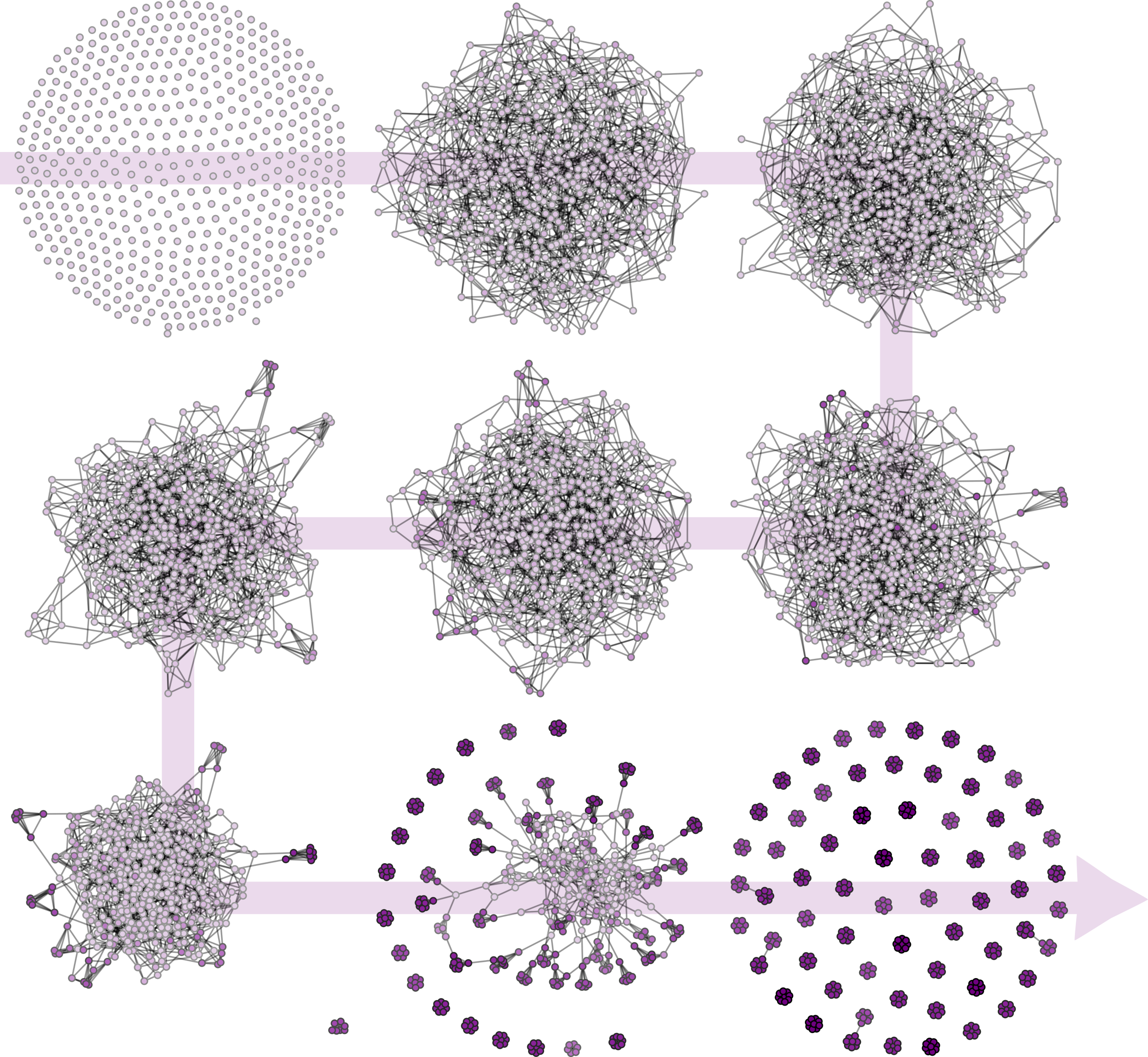}%
\caption{\label{fig:order2freeze} Aggregation kinetics trajectory for the second order model at $T=0.1$; frames show states at 0 and $3125 \times 2^{(0,1,2,3,4,5,6,7)}$ realized events (normalized times $t/A=(0, 1.1, 2.6, 5.9, 12.8, 26.5, 54.6, 136.1, 20785.4)$).  Nodes shaded by triangle membership (darker=more triangles).  A transient gel phase forms quickly, and gradually anneals to the condensed droplet-like phase.}
\end{figure*}

Below their respective critical temperatures, both second and third-order models converge to phases consisting of isolated droplets.  How does this process occur?  Figure~\ref{fig:order2freeze} shows an example of a typical condensation trajectory for the second-order model below the critical temperature ($T=0.1$).  From an initially monomeric state, the system collapses rapidly into a loose (poorly triangulated) gel, which then proceeds to slowly anneal.  Where dense local clusters form, these obtain additional stability by forming additional internal interactions and breaking ties to the rest of the gel (which produce numerous, disfavored, open two-paths).  This gradually leads these increasingly ``droplet-like'' clusters to break off from the gel, forming a state comprised of a mixture of stable droplets and unstable gel.  Eventually, the gel decomposes altogether, leaving only the dense droplets.  These have only very transient interactions, as ties bridging droplets create large numbers of open two-paths; by turns, clusters are constrained from growing to a size much greater than the maximum coordination number (here, 12), since this likewise requires the clusters to sustain high open two-path counts.  

\begin{figure*}[t]
\includegraphics[width=\textwidth]{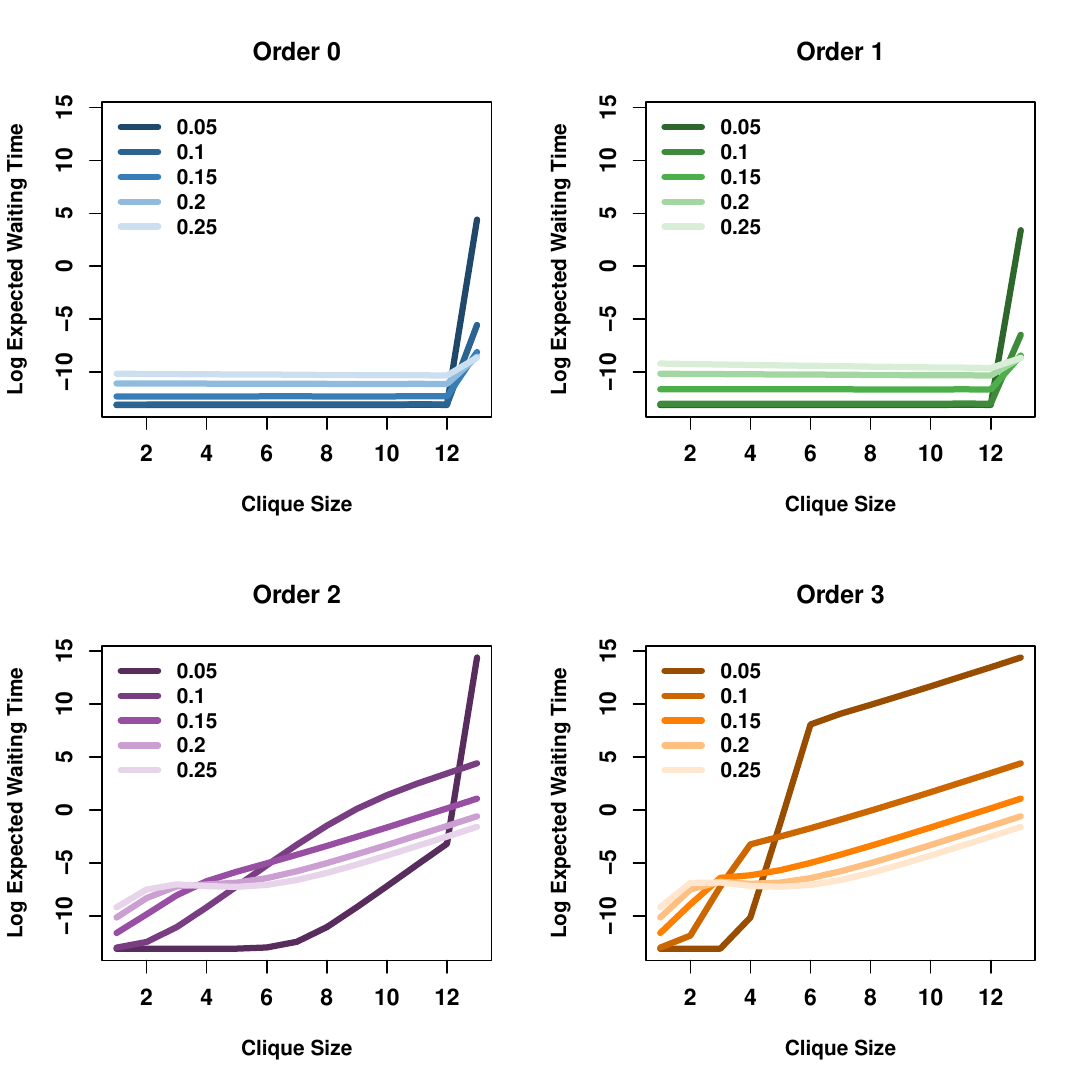}
\caption{\label{fig:wait} Log expected event times (relative to collision rate) for first passage from pure $k$-clique phases, by clique size, model order, and temperature.  Cliques have little stability in low-order models, while high-order models selectively stabilize large cliques at low temperature.}
\end{figure*}

The local stability of droplet-like phases increases sharply at low temperatures, as shown in Figure~\ref{fig:wait}.  Here, we show the log expected waiting time (in collision rate units) for a binding or unbinding event in homogeneous phases consisting of $k$-cliques ($N \approx 1000$).  The zero and first order models show consistently low stability for cliques of all possible orders, with the exception of some increase in stability for the maximum clique size at very low temperature (driven by the inhibition of tie formation for vertices of maximum degree, leaving only dissolution events).  The addition of libration effects alters this picture, as such effects inhibit both formation of cross-clique ties (which produce $2(k-1)$ open 2-paths) and dissolution of within-clique ties (which in turn will produce $k-2$ open 2-paths). This begins to stabilize droplet-like phases, with the increasing favorability of bound interactions at low temperature combining to selectively increase the stability of phases with larger clique sizes (and, eventually, to destabilize small cliques, with the favorability of tie formation overcoming the inhibitory effect of two-path formation).  This dynamic thus favors consolidation, at low temperature, into cliques of maximum size.

The further addition of dihedral effects has no impact on tie dissolution (since no chordless 4-lines can be created by removing an edge from a clique), but does impact tie formation rates.  In particular, addition of an edge between two $k$-cliques creates $(k-1)^2$ chordless 4-lines, an effect which dominates that of libration for $k\ge 4$.  At higher temperatures, cross-clique ties are already sufficiently disfavored that kinetics are driven largely by edge loss, and one thus sees little departure from the second-order model.  However, once well below the critical temperature, this additional inhibition on formation of cross-clique ties combined with increasing tie stability greatly slows kinetics for phases composed of large cliques.  Once formed, droplet-like structures will hence become very stable at low temperature, very slowly annealing towards maximum size; kinetic trapping of mid-sized droplets can be expected at very low temperatures, a pattern that departs from what is seen in the second-order model.

From a physical standpoint, these effects may be interpreted as arising from the fact that while homogeneous, droplet-like states require more local bound interactions than looser structures, they also constrain librational and ``twisting'' motions that can tear aggregates apart.  Here, we model these effects only indirectly, via their coupling to the aggregation graph.  However, their presence notably alters the behavior of the model in the low-temperature regime.

\section{Conclusion} \label{sec:conc}

This paper advances the network Hamiltonian models by incorporating new motional degrees of freedom into the framework. We formulate the description of two-body, three-body, and four-body motional degrees of freedom, and examine how the consideration of these factors influence the topological characteristics of aggregation graphs via simulation.  In particular, we consider aggregation in a highly simplified system containing only a base energy for formation of bound interactions, plus motional effects, allowing us to distinguish the impact of the latter on system behavior.

Our simulation experiments show that, at higher temperatures, effects for bond vibrations and librations respectively reduce the mean degree (hence network density) and decrease component size.  Effects for oscillation about dihedral angles show little additional impact in this regime.  While the zero-order and first-order models smoothly converge to similar behavior in the low-temperature regime, very different behavior is seen for models incorporating higher-order motional effects.  As one approaches $T_c\approx 0.12$ for the librational model, one obtains condensation to a loose (poorly triangulated) gel phase, followed by a sharp transition below the critical temperature to a phase composed of small, highly triangulated, droplet-like clusters.  By contrast, incorporation of dihedral oscillations prevents consolidation into a gel-like phase, instead giving rise to small, primarily linear oligomers that sharply condense to a droplet-like phase at $T_c\approx 0.13$.  

Examination of the kinetics of aggregation behavior suggests that the transition from the gel phase to the droplet phase for the second-order model occurs by annealing of the gel into locally dense clusters that consolidate and bleb off from the main body of the gel.  Once formed, such droplet-like clusters are stabilized both by the unfavorability of bridging ties from the droplet to other structures and by the unfavorability of creating nulls within cliques (both of which create large numbers of open two-paths).  In the case of the third-order model, droplets obviate chordless 4-lines (and hence cliques are favorable where the cost of the requisite number of edges can be overcome).  Once formed, clusters of size $\ge 4$ are kinetically stabilized by the large number of 4-lines created by formation of bridging ties, as well as the same open two-path effects that disfavor tie dissolution within clusters.  This can potentially lead at low temperature to kinetically trapped mixtures of medium to large droplet-like clusters, while the purely librational model shows high stability in the same regime only for clusters of maximum size.

While we consider here only minimal models, the terms employed are easily implemented within a standard ERGM framework, and thus can be used in conjunction with more general classes of NHMs.  It is hoped that these will further broaden the range of molecular systems that can be studied using this class of models, thus complementing other coarse-graining strategies for studying condensation, aggregation, and related phenomena.

\begin{acknowledgments}
    This research was supported by NASA award 80NSSC20K0620, and NIH award 1R01GM144964-01.
\end{acknowledgments}

\section*{Author Contributions}

P.H.: formal analysis; investigation; software; writing - original draft; writing - review and editing

E.M.D.: investigation; visualization; writing - original draft; writing - review and editing 

C.T.B.: formal analysis; investigation; visualization; software; conceptualization; supervision; funding acquisition; writing - original draft; writing - review and editing

\bibliography{ctb}

\end{document}